# High aspect ratio silicon structures by Displacement Talbot lithography and Bosch etching

Konstantins Jefimovs[*a,b], Lucia Romano[a,b,c], Joan Vila-Comamala[a,b], Matias Kagias[a,b], Zhentian Wang[a,b], Li Wang[d], Christian Dais[d], Harun Solak[d], Marco Stampanoni[a,b]

[a]Swiss Light Source, Paul Scherrer Institut, 5232 Villigen PSI, Switzerland; [b]Institute for Biomedical Engineering, University and ETH Zürich, 8092 Zürich, Switzerland; [c]Department of Physics and CNR-IMM- University of Catania, 64 via S. Sofia, Catania, Italy; [d]Eulitha AG, 5416 Kirchdorf, Switzerland

*konstantins.jefimovs@psi.ch

## ABSTRACT

Despite the fact that the resolution of conventional contact/proximity lithography can reach feature sizes down to ~0.5-0.6 micrometers, the accurate control of the linewidth and uniformity becomes already very challenging for gratings with periods in the range of 1-2 µm. This is particularly relevant for the exposure of large areas and wafers thinner than 300 µm. If the wafer or mask surface is not fully flat due to any kind of defects, such as bowing/warpage or remaining topography of the surface in case of overlay exposures, noticeable linewidth variations or complete failure of lithography step will occur.

We utilized the newly developed Displacement Talbot lithography to pattern gratings with equal lines and spaces and periods in the range of 1.0 to 2.4 µm. The exposures in this lithography process do not require contact between the mask and the wafer, which makes it essentially insensitive to surface planarity and enables exposures with very high linewidth uniformity on thin and even slightly deformed wafers. We demonstrated pattern transfer of such exposures into Si substrates by reactive ion etching using the Bosch process. An etching depth of 30 µm or more for the whole range of periods was achieved, which corresponds to very high aspect ratios up to 60:1. The application of the fabricated gratings in phase contrast x-ray imaging is presented.

## 1. INTRODUCTION

Recent development of Displacement Talbot lithography (DTL) [1] proved that this method is very efficient for patterning periodic structures with features ranging from a few micrometers down to sub 100 nanometers. Periodic structures, such as linear gratings or two dimensional arrays of dots that have been printed with DTL, find their application as optical components for photonic crystals, anti-reflection structures, wire-grid polarizers, bio-sensor arrays, plasmonics, spectroscopy etc. While this technology excels in its capability to quickly and reliably print sub-micrometer arrays on a large scale below the resolution of a standard visible and UV lithography, it also shows clear advantages for some lower resolution applications (around one micrometer and above), for which conventional photolithography may also be used. Patterning in this range (1-2 micrometers) by photolithography requires a perfect contact between the mask and the substrate, which means bow and warp free substrates and absence of particles between the contacted surfaces. Yet, even in ideal conditions, the aspect ratio of the structures around one micrometer range becomes limited due to the light diffraction. In comparison, the DTL is essentially a contactless method with theoretically unlimited depth of focus, which solves all the issues mentioned above. Other techniques like steppers or interference lithography would also have such property. However, they would have difficulty with printing gratings without distortion or with exact pitch. An additional advantage of DTL is that it prints gratings with absolute control of pitch and grating phase, defined only by the mask like in case of contact photolithography.

In this work, we demonstrate the potential of DTL technology for the fabrication of gratings for x-ray interferometry [2]. The challenge in x-ray gratings fabrication is twofold. On the one hand, the pitch of the gratings should be relatively small – typically, a few micrometers or even in sub-micrometer range. On the other hand, the lines forming the gratings must have high aspect ratio to provide sufficient phase modulation or absorption of x-rays. We demonstrate the production of gratings with pitches from 1.0 to 2.4 μm exposed by DTL technology and etched into silicon by the Bosch process to depths of more than 30 μm, which corresponds to aspect ratios up to 60:1. Phase gratings with a pitch of 1.3 μm and a height of 24 μm (providing π-phase shift at 17 keV x-ray energy) were used to build a dual phase grating interferometer for x-ray imaging. Examples of the differential phase contrast and dark-field x-ray images are shown.

## 2. DISPLACEMENT TALBOT LITHOGRAPHY

The Talbot effect is an interference based self-imaging phenomenon of a periodic structure (grating) illuminated by a monochromatic and collimated light beam. Self-images repeat at periodic intervals in the light propagation direction. For a linear grating and collimated light illumination, the Talbot self-images appear and repeat approximately with a Talbot period of $2p^2/\lambda$, where p is the grating period and λ is the light wavelength. In addition, Talbot sub-images (also called fractional Talbot images) are sometimes formed at distances corresponding to integer fractions of the Talbot period depending on the properties of the grating. An example of a simulated Talbot pattern is shown in Figure 1 (left). While with very accurate positioning between the mask and the substrate, such Talbot self-images have previously been used to print high resolution patterns [4, 5], the DTL method exploits the periodicity of the Talbot pattern along the light propagation direction. In DTL, the distance (or gap) between the mask and the substrate is changed by one or multiple Talbot periods during the exposure in such a way that average intensity is received by the photoresist, as depicted in Figure 1 (middle). As a result, a total dose profile as shown in the Figure 1 (right) is obtained. The displacement over a Talbot period leads to an exposed pattern independent of the absolute value of the gap between the mask and the substrate. Since no contact between the mask and the substrate is anymore required, the technique is compatible with the use of non-flat and thin substrates, as well as thick photoresists. Moreover, this method is much less sensitive to defects or contaminations on the surface of the resist. Depending on the grating specifications, different periodicities can be obtained in the resulting dose profile. The details of the principle of DTL have been published elsewhere [1, 6].

Figure 2 (top) shows the commercially available PhableR 100 system from Eulitha AG (Switzerland) capable of performing DTL exposures using phase-shifting and amplitude-type (Cr) masks with high uniformity over large areas. The standard system as used in this work can expose wafers up to 100 mm in diameter. Exposures typically take less than one minute depending on the resist sensitivity and mask type (i.e phase-shifting or amplitude). Cross-section and top-down images of various linear gratings with pitches ranging from 2.4 μm down to 0.8 μm are shown in Figure 2 (lower panels). The SEM images are taken after development of photoresist.

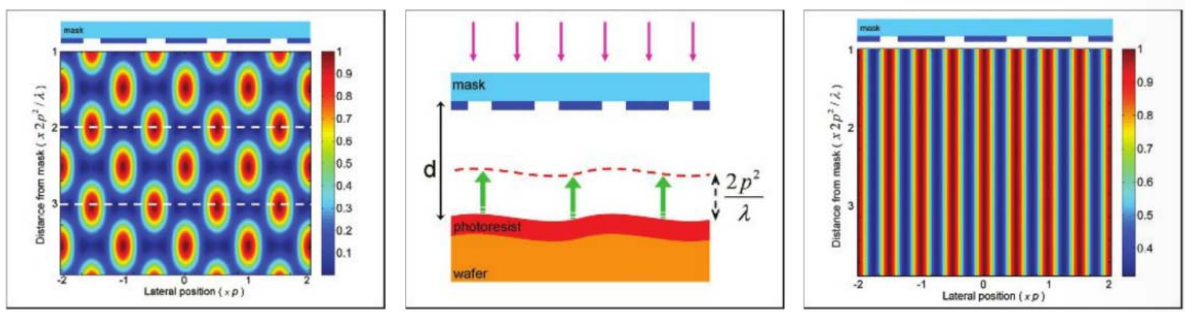

**Figure 1.** A Talbot pattern formed in the vicinity of the mask (left), schematic presentation of the scanning of the substrate during the exposure (middle) and resulting dose profile obtained in the photoresist layer (right). Adapted with permission from Ref. [1].

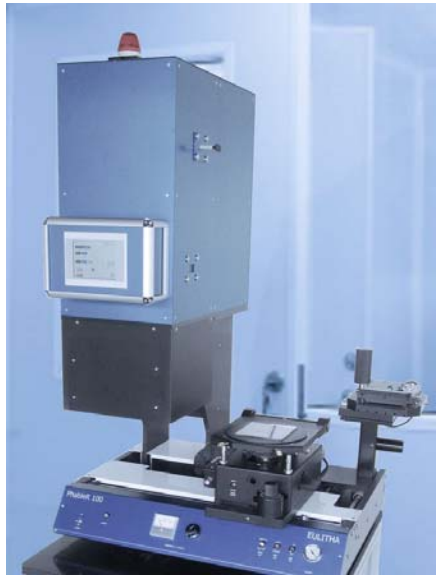

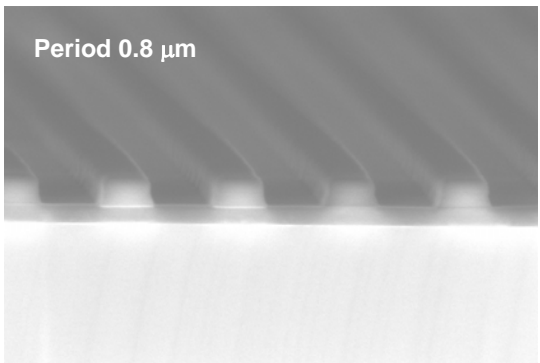
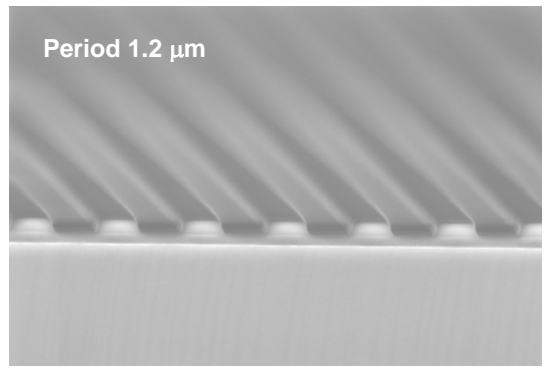
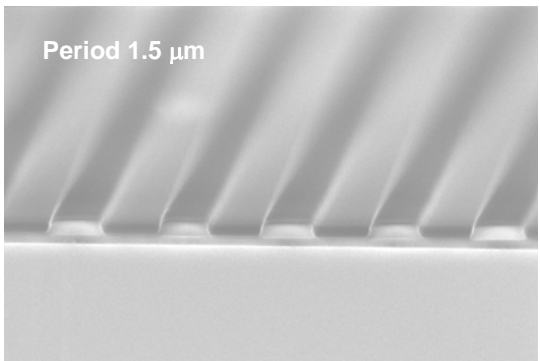
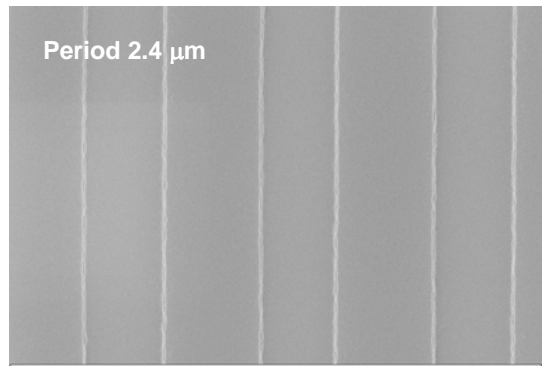

**Figure 2.** A photograph of PhableR100 system from Eulitha AG (top) and SEM images of binary gratings with period ranging from 0.8 to 2.4 μm exposed using PhableR100 system and after photoresist development.

## 3. GRATING FABRICATION

In order to avoid loses due to absorption in the grating based x-rays interferometer setup, the substrates supporting the gratings should be as thin as possible. However, handling of very thin substrates is challenging due to their fragileness. As a compromise, we have chosen 4-inch double side polished silicon wafers with a thickness of 250 micrometers to fabricate the gratings in our case. After the DTL exposure step and the development of the photoresist layer, the gratings were etched into Si substrate as schematically shown in Figure 3. First, the pattern is transferred from photoresist into an underlying antireflective coating by reactive ion etching (RIE) in oxygen plasma. Then, the pattern is further transferred into a Cr hard etching mask by RIE in a $Cl_2$ based process. After Cr etching the residuals of ARC and photoresist layers are removed by RIE in oxygen plasma. Finally, high aspect ratio structures are etched into the Si substrate using $SF_6/C_4F_8$ based Bosch process [7]. The Bosch etching step was performed in a Plasmalab100 system from Oxford Plasma Technologies. A 2-4 mm wide region at the edges of the wafers was covered by a clamping ring, which improves the thermal contact between the wafer and the temperature controlled electrode during the etching process. As a result, we achieved uniform gratings over areas with a diameter of about 90 mm. Gratings with pitches from 1.0 µm to 2.4 µm were etched. The goal for the etching step was to demonstrate the Si structure heights above 30 µm. Achievement of this very high-aspect-ratio and high-resolution structures required careful tuning of all process parameters such as gas flow rates, pressure and RF power. Our results demonstrate the basic capability of the process to achieve this once all the parameters are optimized. Examples of cross-section images taken from the resulting structures are shown in Figure 3.

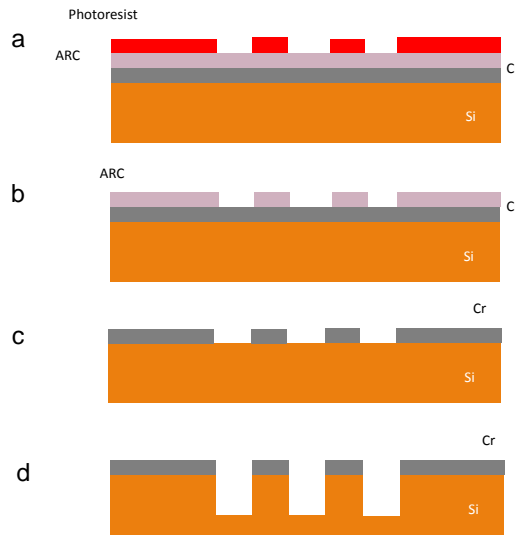

**Figure 3.** Schematics of the fabrication process: a) exposure and development of photoresist; b) dry etching into underlying antireflective coating; c) dry etching into Cr hard mask; d) Bosch etching of Si substrate.

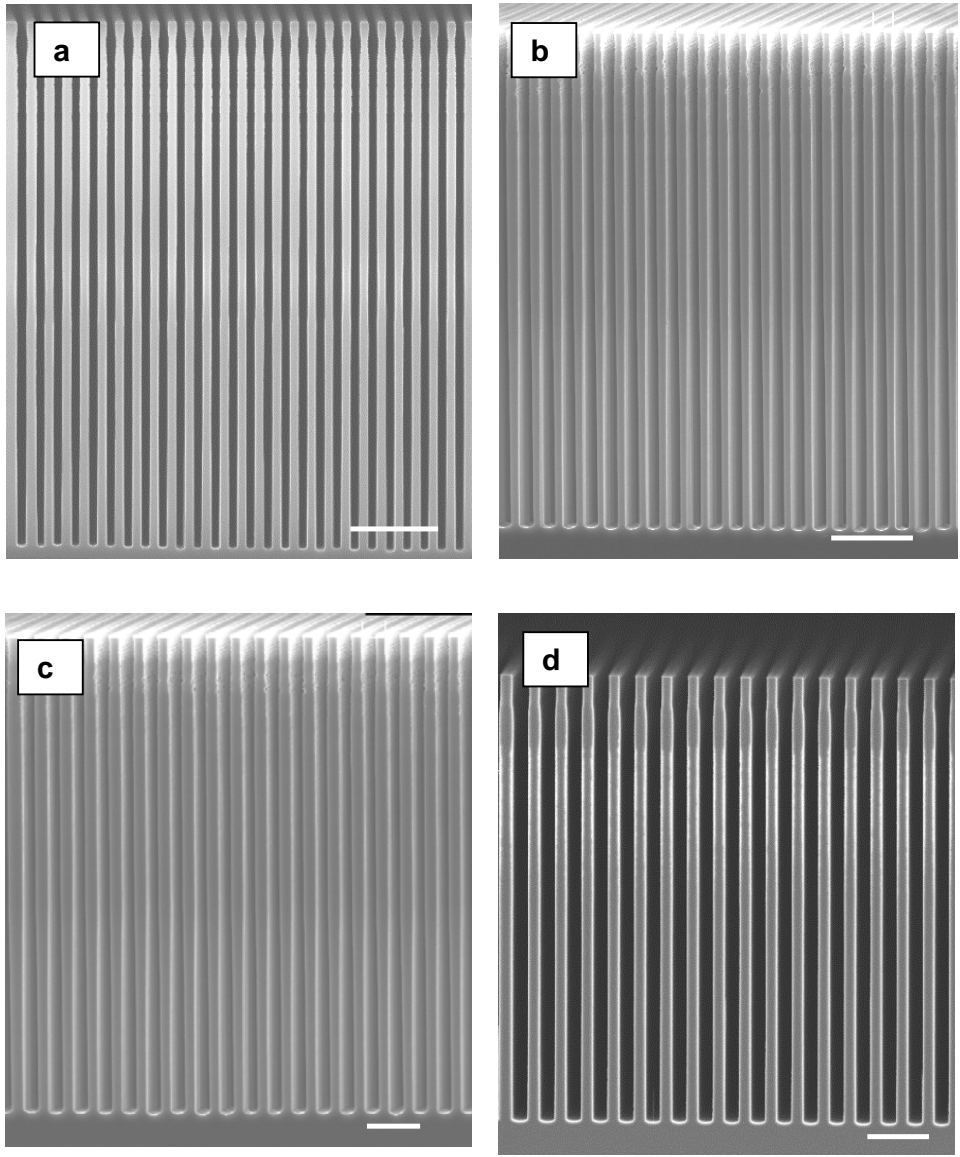

**Figure 4.** Examples of SEM cross-section view of the gratings in Si produced by a combination of displacement Talbot lithography and Bosch process. Grating period (p) and heights (h): a) p=1.0 µm, h=30 µm; b) p=1.5 µm, h=38 µm; c) p=2.0 µm, h=42 µm; d) p=2.4 µm, h=41 µm. Scale bar length in all images is 5 µm.

## 4. X-RAY IMAGING EXPERIMENTS

The fabricated gratings have been successfully applied in a recently developed x-ray interferometer based on two phase shifting gratings that can be used for differential phase contrast as well as dark-field X-ray imaging. A thorough description of the interferometer can be found here [3]. The experimental setup is presented in Figure 5. The first phase grating $G_1$ generates an intensity distribution due to the Talbot effect (which by coincidence is already discussed above in the context of the principle behind the DTL technique). This intensity distribution is then used as a structured illumination for the second phase grating $G_2$. By appropriately choosing the distances between the gratings a large pitch fringe pattern can be generated at the detector plane. When a sample is introduced in the beam (just before the first grating $G_1$) the recorded interference fringes will be affected in three distinct and measurable ways: 1) due to X-ray absorption their average intensity will be reduced; 2) due to refraction their phase will be shifted; and finally 3) due to small X-ray scattering their visibility (modulation) will be reduced as well. Those three signals can be retrieved through Fourier analysis [8].

In order to design a compact system with a large field of view, small grating periods (in the range of 1 μm) are necessary. This is due to the fact that large grating periods lead to large Talbot distances through the relation between the Talbot period and the grating period given above. The effect is quite strong since the distances vary quadratically with the pattern period. We performed imaging experiments to demonstrate the performance of the fabricated gratings. Two identical phase gratings $G_1$ and $G_2$ with a pitch of 1.3 μm and grating lines with a height of 24 μm in silicon corresponding to a π phase shift at 17 keV x-ray energy were fabricated. A photograph of the two 55×75 mm² sized gratings is shown at the bottom right of Figure 5. An SEM image of the grating cross section is also shown in the same figure. As a source we used a microfocal x-ray tube from HAMATSU with a W source size of 9.5 μm (at 70 kVp and 100 mA). This small source size provided enough coherence to generate interference from the phase gratings. The generated interference fringes were recorded with the PI-SCX:4300 detector with a pixel size of 24 μm from Princeton Instruments. The design energy of the system was chosen to be 17 keV, and the distance $l_1$ to 50 cm. Distances, $d_1$ and $l_2$ were chosen as the first fractional Talbot distance at the design energy. By setting $d_2 = d_1$ a symmetrical system design was achieved and $d_1 = d_2$ was approximately 5 mm. The fringe visibility at this configuration was measured to be 20% at the detector plane. Figure 6 shows one example of the acquired x-ray images of a fish. The sample was placed upstream from the gratings. The absorption, differential phase contrast and dark-field (scattering) images were taken during one acquisition.

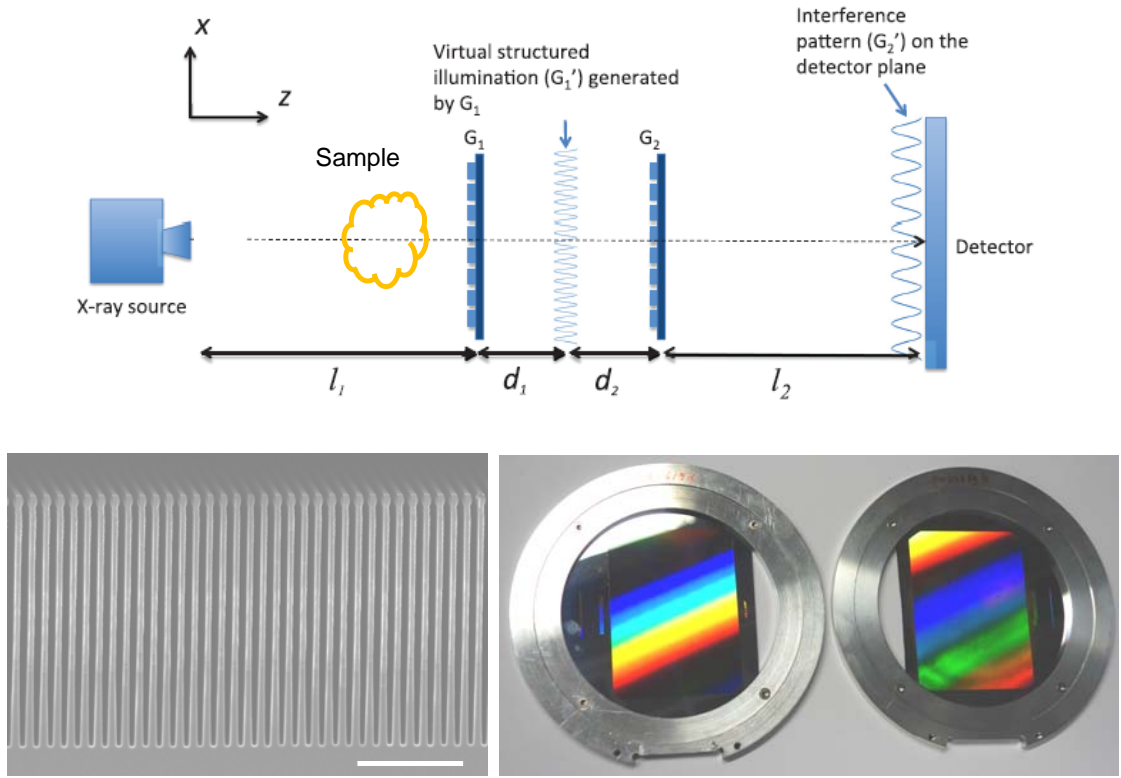

**Figure 5.** Schematic presentation of the measurement setup (top); cross-section SEM image (left bottom) of one of the gratings and photograph (right bottom) of two gratings mounted in grating holder. Both gratings have period of 1.3 μm, lines height of 24 μm, and the size of 55x75 mm². Scale bar length is 10 μm.

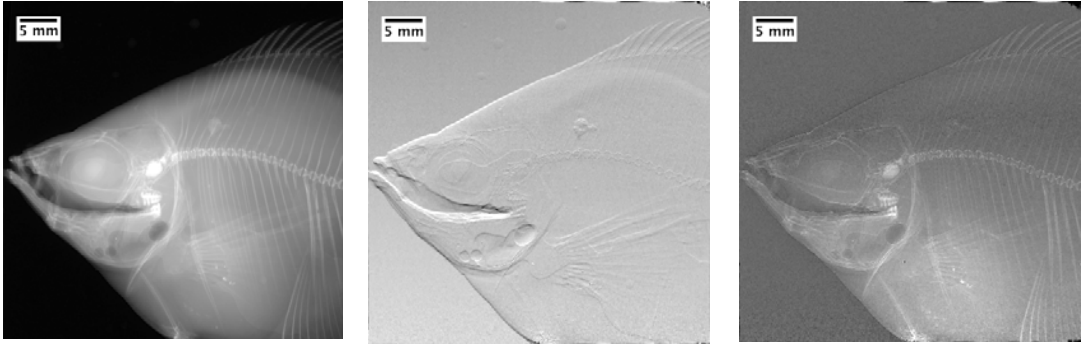

**Figure 6.** Absorption (left), differential phase (center) and dark-field (right) x-ray images of a fish.

## 5. CONCLUSIONS

Our results demonstrate the fabrication of large-area, uniform silicon gratings with pitches in the range of 1.0-2.4 μm at an affordable cost with the use of the Displacement Talbot lithography technology for patterning and the Bosch process for silicon etching. The quality of the printed patterns in terms of uniformity and smoothness are enabling factors in achievement of the required high aspect ratio over large areas. Such gratings are key components that enable construction of compact systems for phase contrast X-ray imaging with a wide-ranging application potential in medicine, biology and material science.

## ACKNOWLEDGEMENTS

Authors would like to thank the staff of TOMCAT group and Laboratory of Micro and Nanotechnology at Paul Scherrer Institut for discussions and technical support. The work was partially funded by the ERC-2012-STG 310005-PhaseX grant, ERC-PoC-2016 727246-MAGIC grant and the Fondazione Araldi Guinetti. We would like to thank M. Bednarzik, C. Wild, D. Marty, V. Guzenko and C. David from PSI-LMN, T. Steigmeier, G. Mikuljan, and C. Arboleda from PSI-TOMCAT for technical support and valuable discussions.

## REFERENCES

[1] H. Solak, C. Dais, and F. Clube, Opt. Express 19, 10686 (2011).
[2] F. Pfeiffer, T. Weitkamp, O. Bunk, and C. David, Nature Phys. 2, 258-261 (2006).
[3] M. Kagias, Z. Wang, K. Jefimovs, and M. Stampanoni, "Dual phase grating interferometer for tunable dark-field sensitivity," Appl. Phys. Lett. 110, 014105 (2017).
[4] D. J. Shir, S. Jeon, H. Liao, M. Highland, D. G. Cahill, M. F. Su, I. F. El-Kady, C. G. Christodoulou, G. R. Bogart, A. V. Hamza, and J. A. Rogers, "Three-dimensional nanofabrication with elastomeric phase masks," J. Phys. Chem. B 111(45), 12945–12958 (2007).
[5] D. C. Flanders, A. M. Hawryluk, and H. I. Smith, "Spatial period division – a new technique for exposing sub-micrometer linewidth periodic and quasi periodic patterns," J. Vac. Sci. Technol. 16(6), 1949–1952 (1979).
[6] H. Solak, C. Dais, F. Clube and L. Wang, "Phase shifting masks in Displacement Talbot Lithography for printing nano-grids and periodic motifs," Microelectron. Eng. 143 (2015) 74–80
[7] I.W. Rangelow, "Critical tasks in high aspect ratio silicon dry etching for microelectromechanical systems," J. Vac. Sci. Technol. A 21 (2003) 1550–1562.
[8] H. Wen, E.E. Bennet, M.M. Hegedus, and S. Rapacchi, "Fourier X-ray scattering radiography yields bone structural information," Radiology 251, 910-918, (2009).